\documentclass[runningheads]{llncs}
\usepackage{graphicx}
\begin{document}
\title{Video Denoising using Channel-Wise Attention Mechanism}
%
%
\author{Aryansh Omray\and
Samyak Jain\and
Utsav Krishnan \and
Pratik Chattopadhyay
}
%
%
\institute{Indian Institute of Technology(BHU), Varanasi, India
\email{\{aryanshomray.cse18,samyakjain.cse18, utsav.krishnan.cse15,pratik.cse\}@iitbhu.ac.in}}
\maketitle              

\def \blocks{Spatio-Temporal Network }

\begin{abstract}
Video Denoising is one of the fundamental tasks of any video processing pipeline. It is different from image denoising due to the temporal aspects of video frames, and any image denoising approach applied to videos will result in flickering. The proposed method makes use of temporal as well as spatial dimensions of video frames as part of a two-stage pipeline. Each stage in the architecture named as Spatio-Temporal Network uses a channel-wise attention mechanism to forward the encoder signal to the decoder side. The Attention Block used in this paper uses soft attention to ranks the filters for better training.

\end{abstract}
\section{Introduction}
Though the advances in the sensor hardware technology have increased the quality of images and videos dramatically over the past few years but the problem of inherent noise in the images and videos seems to be problem. These effects are generally due to low photon count and limitation of the physical imaging sensor present in DSLR, smartphone, medical imaging devices, digital telescopes.\\
Usually, it has been seen that noise is additive in nature, and this property has been proved to be very useful in deep learning work which involves building models that are robust to random noise perturbations. The convolutional neural network described later in this paper makes use of residual connections\cite{DBLP:journals/corr/HeZRS15} to exploit the additive nature of noise. Traditional denoising methods have been subjected to low PSNR values and high run times. Therefore we see that the algorithms employing deep learning are better and can help us achieve real time video denoising. Unlike image denoising, which only deals with the spatial properties of the image whereas video denoising methods also need to look at the temporal effects of the videos. It has been a common practice of using U-net for Gaussian noise denoising.\\

\section{Related Work}
Denoising is one of the fundamental task of any post-processing pipeline for both images and videos. However, the methods for video denoising have been studied less than that of image denoising. Here, we discuss some recent advances in both image and video denoising after a brief discussion on various types of noise and it's different properties.
\subsection{Noise}
Noise refers to the random variation of brightness or color in images/videos during capturing due to sensors present in the camera\cite{noise}. Different types of noise have been studied in literature and will be discussed here. However, the Deep Neural Network presented in this work focuses mainly on Gaussian noise removal from videos.\\
There are some other kind of noise such as Salt \& Pepper Noise\cite{saltandpepper}, Poisson Noise\cite{poissonnoise}, Fractal Noise\cite{hoop2000fluctuations}, Periodic Noise\cite{yoshizawa2011noise}, Speckle Noise\cite{7053819} have been studied. The noise distributions are generally continuous, except some of them being discrete. These noises tend to hamper the visual perception as well as the low-level information present in the image and video signals.\\
However, the real noise which comes from images and videos has a complex mixture of different distributions. The actual type of distributions depends on many factors on conditions in which sensors were used. However, it is seen that low-light conditions often result in more signals than in luminescent conditions due to low-photon count. In addition to this, noise also depends on the quality and type of sensors used. Hence, due to this random nature of real noise, developing methods to remove it poses a difficult challenge. Thus, the development of proper methods is essential to improve the aesthetic quality of an image.\\

\subsection{Image Denoising}
Initial methods in the domain of image denoising deployed spatial filtering techniques such as Min \& Max filters, which targets the least and the most intense regions of a digital image. Min filters serve well in the presence of Salt Noise, whereas Max filters are suitable for Pepper Noise. Similarly, Median filters\cite{median} were developed, which works well for Salt and Pepper Noise. Gaussian filters\cite{Gaussian_filter} make use of the property that Fourier Transform of a Gaussian is also a Gaussian. Using this property, the signal can be transformed with Fast Fourier Transform to perform faster denoising. Among other filter-based methods, Bilateral filters\cite{Bilateral_filter} and Non-local means filter\cite{buades2005non} are also quite popular.\\
A common problem with filter-based denoising methods is that they were suitable for only a few specific types of noise.
Further developments were done to perform image denoising on real noise, and one exciting approach is BM3D\cite{BM3D}, which uses similar 2D image fragments into 3D data arrays based on similarity. Another image denoising that was a variant of the non-local means filter was the Non-local Bayesian Image Denoising\cite{buades2005non} approach, which like BM3D\cite{BM3D} give good results. Nevertheless, still generalization seemed to be a significant issue across all methods. And with the advent of deep learning, it was realized that the denoised results on metrics like PSNR\cite{PSNR} could be improved further by applying deep learning-based denoising methods.\\
Different types of deep convolutional networks such as DnCNN\cite{DnCNN}, FFDNet\cite{FFDNET}, MWCNN\cite{MWCNN} have been developed to perform denoising of images. DnCNN\cite{DnCNN} employs a feed-forward CNN network using residual learning\cite{DBLP:journals/corr/HeZRS15} in addition to batch normalization\cite{bn}, regularization for image denoising. FFDNet\cite{FFDNET} downsamples input images into sub-images before passing into the model for non-linear mapping and upsamples the output into the denoised image. However, most of the above approaches still suffer from unrealistic assumptions, e.g., the noise distribution is Gaussian and the noise present is additive.\\
As an improvement, a few blind image denoising approaches were developed, such as using a generative adversarial network for blind denoising\cite{blind_image_denoising} to extract a noise block of the image and adds it to the original image which is next passed through a convolutional neural network to estimate the noise. This method has been sen to provide encouraging results, but it considers the noise present to be additive. \\
More recent approaches towards image denoising employ U-Net\cite{unet} as their core architecture and have shown promising results. Very recently, few approaches have been developed that consider challenging situations such as denoising in extreme low-light situations\cite{seeing_in_the_dark,our_image_d_paper}. These networks also employ the U-Net as their main architecture for spatial denoising\cite{spatial_denoising} but introduce novelty in the training loss function. For example, in \cite{our_image_d_paper} the loss function is computed from a weighted average of PSNR\cite{PSNR}, SSIM\cite{SSIM}, Edge Loss\cite{canny} and Mean Squared Loss\cite{L2}. Successful application of deep learning in the image denoising task motivates further research in the video denoising domain.\\

\subsection{Video Denoising}
The task of video denoising is different from that of image denoising in the sense that in the former, both spatial and temporal aspects need to be considered. Applying image denoising techniques to videos in a frame-by-frame manner produces severe inconsistencies in resulting frames, which is known as flickering.\\
One of the first approaches to tackle noise in video signals was VBM4D\cite{V-BM4D}, which uses non-local grouping and collaborative filtering by stacking multiple video frames in the $4^{th}$ dimension as compared to VBM3D\cite{VBM3D} which uses the 3D structure for video denoising. This 4D structure is useful in modeling both temporal and spectral correlation to eliminate the effects of flickering. Another similar extension of a non-local Bayesian image denoising algorithm to video denoising is the work in \cite{VNLB}. However, these approaches are capable of handling only specific types of noise and fail to generalize well for different types of noise. Moreover, it suffers from high processing time, making it unsuitable for use in video cameras. Given all these limitations, deep learning has shown a lot better results in temporal denoising and low processing time, making it suitable for real-time denoising purpose.\\
Approaches such as DVDNet\cite{Dvdnet} and FastDVDNet\cite{FastDvdnet} focus on eliminating additive white Gaussian noise(AWGN) by employing residual learning, which leads to faster convergence of neural networks. Here, multiple U-Net\cite{unet} are used for spatial and temporal denoising. While the DVDNet\cite{Dvdnet} algorithm is 25 times slower than \cite{FastDvdnet} due to the use of optical-flow estimation\cite{Deepflowalgorithm} for establishing temporal correlations between frames. Due to the assumption that the noise's nature is additive white gaussian noise, these approaches do not work well for real noise scenarios.\\
Some work involving recurrent denoising auto-encoder\cite{Denoising_Autoencoder} has been done, which assumes Gaussian noise, which is added to a clean image and then the denoising auto-encoder\cite{Denoising_Autoencoder} estimates the clean denoised image using score matching. To maintain temporal stability in denoising auto-encoder\cite{Denoising_AE} LSTM\cite{LSTM} were also used. The architecture employs an encoder-decoder structure in which the subsequent outputs of the encoder side were processed in an LSTM\cite{LSTM} manner to maintain temporal correlations.\\
A few research work on blind denoising has been done to address the noise generalization problem, e.g., ViDeNN\cite{VideNN} for deep blind video denoising and \cite{model_blind} which has been seen to work well for different types of noise. However, we observe that on an average the work in \cite{model_blind} shows a lower PSNR value as compared to approaches purely based on Gaussian additive noise denoising.\\
Much recently, some research has been done on using kernel predicting convolutional neural networks for denoising\cite{KPN,Disney_KPCN}. The network proposed predicts different weights for each pixel, and then the denoised frame is generated. These approaches are generally based on conventional filter-based approaches in image denoising but use deep learning to train those weight vectors while using separate networks for temporal and spatial denoising. Use of an asymmetric loss function\cite{Disney_KPCN} solves the problem associated with this approach.\\
By looking at different advantages and disadvantages of different approaches, we now present our method to perform video denoising, which satisfactorily performs for different types of noise at different levels. Our approach employs a two-layered architecture which follows from FastDVDNet\cite{FastDvdnet} with residual connections\cite{DBLP:journals/corr/HeZRS15} between the mid-frame and the final output. Each \blocks in the architecture is based on U-Net\cite{unet} architecture with the concatenation of filters replaced by a channel-wise attention mechanism.

\section{Dataset}
The training dataset comprises of about 60 videos taken from various sources while maintaining low compression rates and high picture quality across all videos. All videos were digitally compressed in MP4 format while downloading from source and have a resolution of at least $1280\times720$. The length of videos ranges from $10$ seconds to about $200$ seconds, with the majority of videos being about $100$ seconds long. Upon converting the videos into individual frames, the total number of frames in the videos comes close to $180,000$. The videos were captured at about 30 frames/second using high-quality digital cameras.\\
The videos in the training dataset belong to different types of lighting conditions to help the model generalize and give better results. Some of the videos are static, while others have slow-moving elements, while others are fast-moving ones, while others are regular videos.\\
To prepare the dataset for training, Gaussian noise was added into into individual frames of each video from the dataset. The noise was sampled randomly from a Gaussian distribution having $\sigma \: \in \{5,\,10...\,45\}$ and added to each frame during preprocessing.\\
Let $X_i$ denotes the original frame, $X'_i$ denotes the noisy frame and $\mathcal{N}(\mu,\,\sigma^{2})$ denotes Gaussian noise of mean $\mu$ and standard deviation $\sigma$, then\\
\begin{equation}\label{eq:1}  
X'_i\,=\,X_i\,+\,\mathcal{N}(0,\,\sigma^{2})
\end{equation}

\section{Architecture}
The architecture is designed by taking care of both spatial and temporal aspects of the videos and is trained to minimize the chances of flickering. This section discusses the entire architecture in a very detailed manner at each level of the architecture.


\subsection{The Two-stage Pipeline}
The convolutional neural network to denoise videos is designed as a two-stage pipeline, which takes care of the temporal aspects of the videos to reduce flickering instances. The two-stage design of the network is followed by FastDVDnet\cite{FastDvdnet} architecture. However, the network presented in this paper uses residual connection\cite{DBLP:journals/corr/HeZRS15} between the initial mid-input frame and the output frame instead of using it in each denoising block as in FastDVDnet\cite{FastDvdnet}.\\
The use of residual connection\cite{DBLP:journals/corr/HeZRS15} only between the mid-input and the output frame gives the rest of the network the ability to model the noise effectively, leading to faster convergence. This can be explained that the network does not have to train itself to learn features from the video but the distribution of the noise in the training data.\\
For each pass in the network, a total of five frames are taken as input, and a set of three consecutive frames is passed into the \blocks at the first stage. This is done so that the individual \blocks utilizes the temporal aspects of frames and the spatial effects. The \blocks at the first stage share their parameters as this leads to fewer parameters, and hence it saves memory while training the model. It can also be explained in a way that the individual \blocks need to process the input frames similarly, i.e., to extract the temporal and spatial features of the frames and model the noise distribution in the frames, so it makes sense to share the parameters rather than having three individual networks at the first stage. The total number of frames is taken to be such that for each pass in the first stage, the middle frame in the input frames gets processed by each \blocks since it the frame that is needed to be denoised at each pass. The second stage of the network takes the three outputs from the first stage and passes it into another \blocks to get the output. The output is then added with the middle frame using a residual connection\cite{DBLP:journals/corr/HeZRS15} as discussed earlier to obtain the denoised frame.\\


\subsection{\blocks}
The design of \blocks is based on U-Net\cite{unet} architecture except for using Attention-based blocks between corresponding layers in the encoder and decoder instead of concatenating layers from encoder to decoder stage between corresponding layers in the architecture.\\
The Attention Block is added to the corresponding levels between the encoder and the decoder part of the \blocks. Two such Attention Blocks are added into each \blocks. Each \blocks takes as input three frames which are then stacked together in the spectral dimension to form a $9\times W \times H$ shaped tensor and then passed into the \blocks.\\
In each \blocks, the stacked images are passed through several convolutional layers. The network uses a Max-Pooling layer to downsample the input features and uses Deconvolutional Layers as an upsampling mechanism. It should be further noted that after each downsampling and before each upsampling operation in both the encoder and decoder side, the layers are connected using an Attention Block to act as a channel-wise attention mechanism. A Batch Normalization\cite{bn} layer then follows each convolutional layer in the network followed ReLU\cite{relu} activation layer to account for the model's non-linearity except the final layer and the deconvolutional layers. The maximum numbers of filters at any stage is $256$ as compared to $1024$ in the U-Net\cite{unet} Architecture to reduce the number of parameters and reduce the time for training the model. Keeping the number of filters 1024 have negligible impact on the performance of model while increasing the training time and memory requirements.\\


\subsection{Attention Block}
The Attention Block used in the architecture to act as a soft-attention mechanism to guide the gradients in favor of biasing the most informative channel in the feature channels based on Sequence and Excitation Blocks\cite{se_network}. The use of Sequence and Excitation blocks in this architecture helps the network in providing feedback to the layers in the decoder part of the \blocks about the important features while discarding the non-important ones. Since the \blocks does not concatenate encoder features to the decoder side like U-Net\cite{unet} or add them as done in FastDVDnet\cite{FastDvdnet}, the model needs a mechanism to forward the encoder signals to the decoder side for passage of the encoder information to the decoder and thus better feature extraction. The use of Attention Blocks helps in reducing the number of parameters of the network compared to U-Net\cite{unet}.\\
Let $X\in\mathbf{R}^{C\times W\times H},X_{out}\in\mathcal{R}^{C}$ be the input and output to the Attention Block respectively, $W_1\in\mathbf{R}^{C\times C/2}$, $W_2\in\mathbf{R}^{C/2\times C}$ be the weight matrices and $b_1\in\mathbf{R}^{C/2\times1}$, $b_2\in\mathbf{R}^{C\times1}$ be the bias terms for fully-connected layers, $F_{mp}$ is the max-pooling function.In Eq. \ref{eq:3}, $F_{mp}$ is used to for all filters instead for each filter for simplicity . Then
\begin{equation}
    \label{eq:2}
    F_{mp}(X_k)\:=\:\frac{1}{W\times\:H} \sum_{i=0}^{W}\sum_{j=0}^{H}\:X_{k}[i,j]
\end{equation}
\begin{equation}
    \label{eq:3}
    X_{out}\:=\: \delta(W_2\,\times\sigma(W_1\,\times\, F_{mp}(X)\,+\,b_1)\,+\,b_2)
\end{equation}

\section{Implementation Details}
The architecture has been implemented in PyTorch\cite{pytorch} framework in python programming language. For training the model, the total number of the epochs is $100$, and the batch size has set to be one. The optimizer used to minimize the loss function is Adam\cite{adam}, and the learning rate for training the model is kept to be $10\textsuperscript{-3}$ for the first fifty epochs and $10\textsuperscript{-4}$ for the rest of the fifty epochs. The initialization of the weights of the networks using Xavier\cite{xavier} initialization using the PyTorch\cite{pytorch} default parameter initialization. The images extracted from the videos are randomly cropped into the size of $512\times512$ to introduce some data augmentation while avoiding overfitting. We trained our model on an Nvidia V100 GPU with 16 GB of memory.

\section{Conclusion}
This paper uses a channel-wise attention mechanism to address the problem of video denoising using a U-Net based auto-encoder Network in a temporal setting. The network tends to minimize the effects of flickering by exploiting temporal nature of videos by using multiple frames in each forward propagation.

%
%
%
%
\bibliographystyle{splncs04}
\scriptsize
\bibliography{ref}
\end{document}